\documentclass{article}
\begin{document}
\title{A Scheme to Classify Topological Property of Band Insulator Based On One-band $U(1)$ Chern Number}
\author{Wu Yidong\\
        Department of Applied Physics,Yanshan University}
\maketitle
\begin{abstract}Topological insulator(TI) is a phase of matter discovered recently.$^{1-3}$ Kane and Mele proposed this phase is distinguished from the ordinary band insulator by a $Z_2$ topological invariant.$^2$ Several authors have try to related this $Z_2$ invariant to Chern numbers.$^{4,5}$Roy find a way to calculate $Z_2$ by Chern Number of one of the two degenerate Bands or one-band Chern number(OBChN).$^5$ However, he give no concrete concrete proof of the equivalence of his $Z_2$ and the $Z_2$ in ref[2] beside ``from the topological considerations of K theory".$^5$ So the importance of OBChN hasn't been recognized by the community. In this letter we prove OBChN determines the $Z_2$ in ref[2]. Then we illustrate OBChN is not only an useful tool to identify TI but also a natural criterion to classify topological property of all time-reversal invariant band insulators.More importantly we find a field in three dimensional TI can be identified with magnetic field with magnetic monopole.\\
\end{abstract}

 We consider a two-dimensional time-reversal invariant tight-binding model. In momentum space the Hamilton take the form
\begin{equation}
   H=\sum_{\mathbf{k}}\mathcal{H}(\mathbf{k})
\end{equation}
 $\mathcal{H}(\mathbf{k})$ is smooth and period in the two dimensional momentum space.
 \begin{equation}\mathcal{H}(k_x+2\pi,k_y)=\mathcal{H}(k_x,k_y)\end{equation}
 \begin{equation}\mathcal{H}(k_x,k_y+2\pi)=\mathcal{H}(k_x,k_y)\end{equation}
 So the Brillouin zone can be considered as a torus $T^2$. Time-reversal invariant symmetry force $\mathcal{H}(\mathbf{k})$ to satisfy
 \begin{equation}\Theta \mathcal{H}(\mathbf{k}) \Theta^{-1}=\mathcal{H}(\mathbf{-k})\end{equation}
 $\Theta $ is the time reversal operator. For simplicity we consider only one pair of degenerate bands. For a given $\mathbf{k}$ the eigenvector of $\mathcal{H}(\mathbf{k})$ expand a two dimensional vector space $V(\mathbf{k})$. All the $V(\mathbf{k})$ form a vector bundle with a torus as a base space.Because of the time-reversal invariant symmetry Chern number of  the associate $U(2)$ principle bundle of this vector bundle is zero.So the vector bundle is a trivial bundle, that is it's homeomorphic to $T^2\times V(\mathbf{k})$.Then we can always find two smooth orthonormal vectors as functions of
 $\mathbf{k}$:$|X^{I}(\mathbf{k})\rangle$ and $|X^{II}(\mathbf{k})\rangle$. For a given $\mathbf{k}$ these two vectors span the vector space $V(\mathbf{k})$. Fu and Kane define the $Z_2$ invariant as the parity of$^6$
 \begin{equation}\Delta=P_{\theta}(\pi)-P_{\theta}(0)\end{equation}
 \begin{equation}P_{\theta}(k_y)=P^{I}(k_y)-P^{II}(k_y)\end{equation}
 \begin{equation}P^{s}(ky)=\frac{1}{2\pi}\int_{-\pi}^{\pi}A_x^s(kx,ky)\mathrm{d}k_x\end{equation}
 where $s=I,II$ $
 A^s_x=i\langle X^{s}|\frac{\partial}{\partial k_x}| X^{s}\rangle$. Notice that we use a different notation from ref[6], where they use $k$,$t$,$T$ we use $k_x$,$k_y$ and $2\pi$ respectively. In ref[6] they prove $\Delta$ is an gauge independent integer,which can be calculated by any pair of globe smooth orthonormal functions. So in the definition of $A^s_x$ we use $| X^{s}(\mathbf{k})\rangle$.
 Now we build a relation between $\Delta$ and the one-band Chern number. If the eigenfunctions of the Hamilton can be defined continuously in the upper half Brillouin zone( $0\leq k_y \leq \pi$), the proof will become almost trivial: by using Stokes theorem it can be show $\Delta$ is just one of the one-band Chern Number. Unfortunately the eigenfunctions cannot be defined continuously in general. So we try to diagonalize the Hamilton in the vector space $V(\mathbf{k})$ and find why the eigenfunctions fail to be continuous. Then we show the discontinuity doesn't affect the relationship between one-band Chern number and the $Z_2$.\\

 Using $|X^{I}(\mathbf{k})\rangle$ and $|X^{II}(\mathbf{k})\rangle$ as the base of $V(\mathbf{k})$ Bloch Hamilton $\mathcal{H}(\mathbf{k})$ is a $2\times2$ Hermite matrix. The matrix element $h_{mn}(\mathbf{k})=\langle X^{m}(\mathbf{k})|\mathcal{H}(\mathbf{k})| X^{n}(\mathbf{k})\rangle$, where $m,n=I,II$. Because of the Smoothness
 of $|X^{s}(\mathbf{k})\rangle$ and $\mathcal{H}(\mathbf{k})$, $h_{mn}(\mathbf{k})$ is smooth too. Hermitianity of the matrix means $h_{11}$ and $h_{22}$ is real and $h_{12}=h_{21}^*$. The eigenvalue of the matrix is
 \begin{equation}E_{1,2}(\mathbf{k})=\frac{(h_{11}+h_{22})\pm \sqrt{(h_{11}-h_{22})^2+4|h_{12}|^2}}{2}\end{equation}
 where $1$ corresponds to $+$,$2$ to $-$,they are the up and down branch of the solution.Becourse of the smoothness of Hamilton we can always construct from them two smooth solutions $E^{I,II}(\mathbf{k})$.At the degenerate points of Hamilton we can take the limit of solution so the eigenvectors $u^{I,II}(\mathbf{k})$ can locally defined in general.$^{19}$. The orthonormal eigenvectors can only be determined up to a $U(1)$ phase factor. In solving eigenvector$|u^{I}(\mathbf{k})\rangle$(correspond to $E^{I}$) we choose the first component nonnegative real number and $|u^{II}(\mathbf{k})\rangle$ second component nonnegative real. So we eigenfunctions can be solved as
\begin{equation}
\left(
\begin{array}{ccc}
\langle X^{I}(\mathbf{k})| u^{I}(\mathbf{k})\rangle & \langle X^{I}(\mathbf{k})| u^{II}(\mathbf{k})\rangle  \\
 \langle X^{II}(\mathbf{k})| u^{I}(\mathbf{k})\rangle & \langle X^{II}(\mathbf{k})| u^{I}(\mathbf{k})\rangle
\end{array}
\right)\\
=\left(
\begin{array}{ccc}
 1/\sqrt{1+|c(\mathbf{k})|^2} & -c^*(\mathbf{k})/\sqrt{1+|c(\mathbf{k})|^2}  \\
c(\mathbf{k})/\sqrt{1+|c(\mathbf{k})|^2} & 1/\sqrt{1+|c(\mathbf{k})|^2}
\end{array}
\right)
\end{equation}
where $c(\mathbf{k})=(E^1-h_{11})/h_{12}$ when $u^{I}(\mathbf{k})$ is on up branch,$c(\mathbf{k})=(E^2-h_{11})/h_{12}$ when $u^{I}(\mathbf{k})$  on down branch. At the degenerate point $c(\mathbf{k})$ take the limit,it can be easily shown $c(\mathbf{k})$ is continuous except it become infinite. From the result we see clearly the eigenfunctions are smooth as long as $c(\mathbf{k})$ is finite.\\

First we show the eigenfunctions can be smoothly defined at $k_{y}=0$ and $k_{y}=\pi$ where the base space can be viewed as circles. If at some point or segment $c(\mathbf{k})$ becomes infinite we an always cover it by a slightly larger segment.On this segment we take a different phase fixation as follow
\begin{equation}
\left(
\begin{array}{ccc}
\langle X^{I}(\mathbf{k})| u^{I}(\mathbf{k})\rangle & \langle X^{I}(\mathbf{k})| u^{II}(\mathbf{k})\rangle  \\
 \langle X^{II}(\mathbf{k})| u^{I}(\mathbf{k})\rangle & \langle X^{II}(\mathbf{k})| u^{I}(\mathbf{k})\rangle
\end{array}
\right)\\
=\left(
\begin{array}{ccc}
c'(\mathbf{k})/\sqrt{1+|c'(\mathbf{k})|^2} & -1/\sqrt{1+|c'(\mathbf{k})|^2}\\
1/\sqrt{1+|c'(\mathbf{k})|^2}& c'^{*}(\mathbf{k})/\sqrt{1+|c'(\mathbf{k})|^2}
\end{array}
\right)
\end{equation}

where $c'(\mathbf{k})=c(\mathbf{k})^{-1}$. Because of continuity $c'(\mathbf{k})$ should be finite in the adjacent infinite point of $c(\mathbf{k})$.So the eigenfunctions are smoothly defined on the covering segment. On the boundary point of the covering segment there will be a phase mismatch between two solutions and the phase difference are opposite for the two bands. However we can always multiply a well chosen smooth factor $e^{i\theta(k_x)}$ to $| u^{I}(\mathbf{k})$ and $e^{-i\theta(k_x)}$ to $|u^{II}(\mathbf{k})$ from both side of boundary point to cancel the phase mismatch and joint the eigenfunctions smoothly. Mathematically this process is to find a smooth $U(2)$ transformation to diagonalize the Hamilton on a circle. With our choice the determinant of the $U(2)$ matrix is always equal to $1$.So on the circle Hamilton can be diagonalized by a smooth $SU(2)$ transformation. The fundamental group $\pi_1(SU(2))=\{e\}$,$^7$so the two smooth $SU(2)$ transformation on the boundary circle can be jointed by a smooth $SU(2)$ transformation on the upper half Brillouin zone.With this transformation a new pair of smooth base vectors on the upper half Brillouin zone are defined,we still denote them $|X^{I}(\mathbf{k})\rangle$ and $|X^{II}(\mathbf{k})\rangle$.\\

With the new bases the Hamiltons are diagonalized on boundaries.So $c(\mathbf{k})s$ are zeros on the boundary, by continuity the possible infinite $c(\mathbf{k})$ points are expelled to the middle of upper half Brillouin zone( $0< k_y <\pi$), when it happens on some points, lines or regions we can always cover them by slightly larger region(make sure different covering regions don't overlap) and use the other choice to find the smooth eigenfunctions on the covering regions. Thus smooth eigenfunctions are all well defined on the covering regions and the rest of upper half Brillouin zone(though phase mismatchs present on boundaries),so we can calculate Berry's connection one-form and Berry's curvature two-form on them. \begin{equation}\mathcal{A}^s=\mathbf{A^s}\cdot\mathrm{d}\mathbf{k}\end{equation}
\begin{equation}\mathbf{A}^s=i\langle u^{s}(\mathbf{k})|\bigtriangledown_{\mathbf{k}}| u^{s}(\mathbf{k})\rangle\end{equation}
\begin{equation}\mathcal{F}^s(\mathbf{k})=F^{s}(\mathbf{k})\mathrm{d}k_x\wedge \mathrm{d}k_y\end{equation}
\begin{equation}F^{s}(\mathbf{k})=i( \frac{\partial\langle u^{s}(\mathbf{k})}{\partial k_x}|\frac{\partial| u^{s}(\mathbf{k})\rangle}{\partial k_y}-\frac{\partial\langle u^{s}(\mathbf{k})}{\partial k_y}|\frac{\partial| u^{s}(\mathbf{k})\rangle}{\partial k_x} )\end{equation}

$F^{s}(\mathbf{k})$ is $U(1)$ gauge invariant,so it's independent of the $|u^{I}(\mathbf{k})\rangle$ we choose to calculate it and it's smoothly defined on the whole Brillouin zone.Now we integrate the curvature two form in the upper half Brillouin zone and use Stoke's Theorem on every region.
\begin{equation}\int_{\Omega}F^{s}(\mathbf{k})\mathrm{d}k_x\mathrm{d}k_y=\int_{-\pi}^{\pi}A_x^s(kx,0)\mathrm{d}k_x-\int_{-\pi}^{\pi}A_x^s(kx,\pi)
\mathrm{d}k_x+\sum_j\int_{\partial\Omega_j} \mathbf{A^s}\cdot\mathrm{d}\mathbf{k}\end{equation}

$\Omega$ is the upper half Brillouin zone, $\partial\Omega_j$ is the boundary of the $j-th$ covering region $\Omega_j$. On the $\partial\Omega_j$ there are two line integrals with opposite direction, they don't cancel because the phase mismatch of the eigenfunctions on boundaries. Denotes the phase difference by $e^{i\chi^s_j(\mathbf{k})}$.It can be shown the sum of the two line integration is$^8$
\begin{equation}\int_{\partial\Omega_j}\bigtriangledown_{\mathbf{k}}\chi^s_j(\mathbf{k})\cdot\mathrm{d}\mathbf{k}=2\pi n^s_j \end{equation}

$n^s_j$ is $U(1)$ winding number of $e^{i\chi^s_j(\mathbf{k})}$ on $\partial\Omega_j$,it's an integer.As discussed on the circle the phase mismatch of the two bands is always opposite,so we have $\chi^I_j(\mathbf{k})=-\chi^{II}_j(\mathbf{k})$ ,thus $n^I_j=-n^{II}_j$. Finally $\Delta$ can be expressed as

\begin{equation}\Delta=\frac{1}{2\pi}(\int_{\Omega}F^{II}(\mathbf{k})\mathrm{d}k_x\mathrm{d}k_y-\int_{\Omega}F^{I}(\mathbf{k})\mathrm{d}k_x\mathrm{d}k_y)+\sum_j n^I_j -\sum_j n^{II}_j\end{equation}
It has be shown that $F^{I}(\mathbf{k})=-F^{II}(-\mathbf{k})$. So
\begin{equation}\Delta=\frac{1}{2\pi}\int_{\Omega}(F^{II}(\mathbf{k})+F^{II}(-\mathbf{k}))\mathrm{d}k_x\mathrm{d}k_y+2\sum_j n^I_j=\frac{1}{2\pi}\int_{B}F^{II}(\mathbf{k})\mathrm{d}k_x\mathrm{d}k_y+2\sum_j n^I_j=Ch^{II}+2\sum_j n^I_j
\end{equation}
Where $B$ denote the Brillouin zone.$Ch^{II}$is the Chern number of band $II$.From the last expression we can see $Z_2$ invariant is just parity of the one-band Chern number(OBChN). If several pairs of bands present $Z_2$ is the parity of the sum of OBChNs for each pair\\

With our conclusion TI is identified by odd-parity OBChN or sum of OBChNs,for instance in ref[1-3] the OBChN of occupied bands is $1(-1)$. Despite the fragileness of their edge states the $Z_2$-trivial spin Hall insulator in ref[9] is also gain status in our scheme with a nonzero OBChN equal to $2(-2)$.In the first case there are two edge state on each edge and four in the second. So it suggest the OBChN may provide information about number of edge states.\\

In three dimensional topological indices defined in ref[10] is the parity of OBChN  on three k-space primitive cell(primitive-reciprocal-lattice-vector spanned parallelepiped with origin as center)surfaces and OBChN difference between the cell surface and the parallel middle cross section .$^{11}$ So the strong and weak TI can be easily classified in our scheme. However there are still some topological nontrivial insulators with even OBChNs don't belong to them.$^{12,13}$\\

 Most of first-principle-based researches on TI depend on $Z_2$ calculation or band inversion paradigm.$^{14-17}$ $Z_2$ can be more conveniently calculated through OBChN which can be determined by locally continuous eigenfunctions. Band inversions can be indicated by OBChN change. If we tune some parameter smoothly, Hamilton on a surface will change smoothly too. So with the new parameter we can define a Berry's one-form and $U(1)$ gauge independent two-form on one band in three dimension. Using Stoke's Theorem and $d^2=0$,We find OBChN won't change if Berry's two-form are defined smoothly. So if there is a OBChN change there ought be some singular points of the two-form like singular point of field strength where magnetic charge(monopole)  present in $U(1)$ gauge theory of electromagnetism. The ``topological charge" has a integer multiply $2\pi$ quantity and determine where the band inversion or collision happens. For example in three dimensional k-space primitive cell, the middle cross section can be smoothly translated to cell surface,if there is a OBChN difference between two surfaces some ``topological charge" must present in the half cell.In this way strong TI can be elegantly defined by the  "topological charge" contained in half Brillouin zone.\\

 More interestingly and importantly we find the above mentioned Berry's form in three dimension can be identified with magnetic-monopole-generate magnetic field :one-form corresponding to vector potential and two-form to field strength.$^{18}$ The difference is Berry's form has a more complicate boundary condition(cross sections must be identified with torus).Above mentioned formulation determines the geometry part of the Berry's form(field)theory, the dynamic part can be obtained from Hamilton.$^7$ We hope our work will stimulate the study of magnetic monopole through the Berry's form in three dimensional TI.$^{18}$\\

 In general we suggest all the topological nontrivial time-reversal invariant band insulators may be obtained from topologically trivial(all OBChNs are zeros on all k-space surface it can be defined)band insulators by smoothly introduce some interaction e.g. spin-orbit interaction. During the process bands may invert or collide and generate the topological non-triviality. OBChN on k-space surfaces is guidepost of the process. So OBChNs or ``topological charge" is a natural criterion to classify topological property of time-reversal invariant band insulators.
\begin{itemize}

  \item[1.] C. L. Kane and E. J. Mele, Phys. Rev. Lett. 95, 146802 (2005).
  \item[2.] C. L. Kane and E. J. Mele, Phys. Rev. Lett. 95, 226801 (2005).
  \item[3.]B. A. Bernevig, T. L. Hughes, and S.-C. Zhang, Science
314, 1757 (2006).
  \item[4.]J. E. Moore and L. Balents, Phys. Rev. B 75, 121306
(2007).
  \item[5.]R. Roy, Phys. Rev. B 79, 195321 (2009). In our proof Hermitic of the Hamilton play an important role, so ``the topological considerations of K theory" seems not sufficient.
  \item[6.]L. Fu and C. L. Kane, Phys. Rev. B 74, 195312 (2006).
  \item[7.]M. Nakahara, Geometry, Topology and Physics (CRC, Florida,
2003).
  \item[8.]M. Kohmoto, Ann. Phys. 160, 343 (1985).
  \item[9.]X. L. Qi, Y. S. Wu, and S. C. Zhang, Phys. Rev. B 74,
085308 (2006).
  \item[10.]L. Fu, C. L. Kane, and E. J. Mele, Phys. Rev. Lett. 98,
106803 (2007).
  \item[11.]R. Roy, Phys. Rev. B 79, 195322 (2009).
  \item[12.]S. Murakami, N. Nagaosa, and S. C. Zhang, Phys. Rev.
Lett. 93, 156804 (2004).
  \item[13.]M. Onoda and N. Nagaosa, Phys. Rev. Lett. 95, 106601
(2005).

  \item[14.]S. Chadov et al., Nature Mater. 9, 541 (2010).
  \item[15.]H. Lin et al., Nature Mater. 9, 546 (2010).
  \item[16.]Yan Sun et al.,Phys. Rev. Lett. 105, 216406 (2010)
  \item[17.]W. Feng1, D. Xiao, J. Ding, and Y. Yao,Phys. Rev. Lett. 106, 016402 (2011)
  \item[18.]details to be published soon.
  \item[19.]the singular point that eigenfunctions cannot locally continuously is ``topological charge" .   

\end{itemize}
\end{document}